\documentclass[%
            10pt,%
            aps,%
            prd,%
            reprint,%
            onecolumn,%
            floatfix,%
            nofootinbib,%
            superscriptaddress,%
            notitlepage
            ]{revtex4-2}
\usepackage{graphicx}
\usepackage[export]{adjustbox}
\usepackage{dcolumn}
\usepackage{%
        amsmath,%
        amsfonts,%
        bm
        }
\allowdisplaybreaks
\usepackage{physics}
\usepackage{slashed}
\usepackage[%
        format=hang,%
        singlelinecheck=off,%
        justification=raggedright,%
        caption=false
        ]{subfig}
\usepackage[usenames,dvipsnames,svgnames]{xcolor}
\usepackage{colortbl}
\definecolor{Gray}{gray}{0.75}
\usepackage{tikz}
    \usetikzlibrary{calc}
\usepackage{pgfplots}
    \pgfplotsset{compat=1.18}
\usepackage{multirow}
\usepackage{tasks}
\usepackage{makecell}
\usepackage{hhline}
\usepackage{mathrsfs}
\usepackage[%
        shortlabels,%
        inline
        ]{enumitem}
\usepackage{hyperref}
\usepackage[%
        capitalise
        ]{cleveref}
\crefmultiformat{figure}{Figs.~#2#1\xdef\mycreffirstarg{#1}#3}%
 { and~#2{\crefstripprefix{\mycreffirstarg}{#1}}#3}{, #2#1#3}{ and~#2#1#3}
\PassOptionsToPackage{demo}{graphicx}

\hypersetup{%
        colorlinks=true,%
        linktocpage=true,%
        pdfstartpage=1,%
        pdfstartview=FitV,%
        breaklinks=true,%
        pdfpagemode=UseNone,%
        pageanchor=true,%
        pdfpagemode=UseOutlines,%
        plainpages=false,%
        bookmarksnumbered,%
        bookmarksopen=true,%
        bookmarksopenlevel=1,%
        hypertexnames=true,%
        pdfhighlight=/O,%
        urlcolor=Cerulean,%
        linkcolor=Cerulean,%
        citecolor=Cerulean,%
}

\begin{document}
\title{Feebly-interacting dark matter}
\author{G. Bélanger}
\email{belanger@lapth.cnrs.fr}
\affiliation{LAPTh, CNRS, USMB, 9 Chemin de Bellevue,
             Annecy 74940, France}
\author{S. Chakraborti}
\email{sreemanti.chakraborti@durham.co.uk}
\affiliation{IPPP, Department of Physics, Durham University,
             Durham DH1 3LE, United Kingdom}
\author{A. Pukhov}
\email{alexander.pukhov@gmail.com}
\affiliation{Skobeltsyn Institute of Nuclear Physics, Moscow State University,\\
\hskip5pt     Moscow 119992, Russia}

\keywords{Dark matter, Feeble interactions, Freeze-in}
%

%
\begin{abstract}
We briefly review scenarios with feebly interacting particles  (FIMPs) as dark matter candidates. The discussion covers issues with dark matter production in the early universe as well as signatures of FIMPs at the high energy and high intensity frontier as well as in astroparticle and  cosmology.
\end{abstract}

\maketitle

%
\section{Introduction}

Despite the strong evidence for dark matter (DM) on galactic, clusters and cosmological scales, the nature of  DM remains a mystery. Although the relic density of DM has been measured with great precision $\Omega h^2=0.1188\pm 0.001$~\cite{Planck:2015fie}, this still leaves several possibilities for DM that span dozens of orders of magnitude in mass and/or in interaction strength. The lack of evidence for the well-motivated weakly interacting massive particles (WIMP) in direct/indirect and collider searches aroused the interest 
in opening possibilities  for DM candidates of  different scales and interaction strengths. 
 In this presentation, we will address the possibility  of feebly interacting particles as dark matter (FIMPs) and of  DM production through the freeze-in (FI) mechanism. 
We will discuss the signatures of FIMPs at colliders, in direct detection and in cosmology. 

\section{Dark matter formation}

When DM particles (hereafter denoted as $\chi$) are feebly interacting, they cannot reach thermal equilibrium in the early universe. DM production is achieved through 
the freeze-in mechanism and results from the scattering or decay of SM or BSM particles  in the thermal bath~\cite{Hall:2009bx,McDonald:2001vt}.

The interactions between the dark and the visible sector can be either through non-renormalisable operators or renormalisable interactions. In the former case, known as UltraViolet (UV) freeze-in~\cite{Elahi:2014fsa}, the DM abundance is obtained via effective operators. Here, the renormalisation scale suppresses the interaction strength between the SM and the dark sector and therefore keeps DM out of thermal equilibrium, moreover the DM abundance features  a strong dependence on the reheating temperature. The latter case is known as InfraRed (IR) freeze-in~\cite{Hall:2009bx}, where the freeze-in is generated in a class of models where different sectors are connected via renormalisable operators. A review of FIMP models can be found in  Ref.~\cite{Bernal:2017kxu}.

The number density of DM, $n_\chi$, follows a Boltzmann equation
\begin{equation}
\frac{dn_\chi}{dt}+3Hn_\chi = \sum_X \langle \sigma v \rangle_{X\bar{X}\to \chi\bar\chi} \bar{n}_X^2(T) +  \sum_{X'}\Gamma_{X'\to \chi\bar\chi}(T) \bar{n}_{X'}(T) 
\end{equation}
where $\bar{n}_{X}$ is the equilibrium density, $\langle \sigma v \rangle$ is the cross section for the production of   DM through pair annihilation, $X\bar{X}\to \chi\bar\chi$,   and $\Gamma_{X'\to \chi\bar\chi}$ is the decay width of $X'\to\chi\bar\chi$.  Here $X,X'$ are particles that belong to  the thermal bath. DM is produced from annihilation or decays until the number density of SM particles becomes Boltzmann suppressed, at this point $n_\chi/s$ becomes constant and "freezes-in". This occurs at $T \approx M$ where $M$ is the mass of  the mediator ($X'$) or of DM. 
Note that the term that corresponds to depletion of DM through pair annihilation into SM particles that appears in the evolution equation for the number density in the  freeze-out (FO)  scenario is neglected considering the small initial abundance of $\chi$.

The above equation is usually written in terms of the abundance, $Y_\chi=n_\chi/s$ where $s$ is the entropy density.  The  abundance today ($Y_\chi^0$) is obtained by solving the evolution equation from the temperature at which DM production starts, referred to as the reheating temperature $T_R$, to the present day temperature. In contrast with the case of WIMP freeze-out,  the DM abundance increases with the interaction strength. The typical value of the coupling for weak scale FIMPs is in the range $10^{-10}-10^{-12}$.  Note that when decay is possible it usually dominates over annihilation.  The DM relic density is simply related to $Y_\chi^0$, 
\begin{equation}
\Omega_\chi^{FI} h^2=\frac{m_\chi Y_\chi^0 s_0\,h^2}{\rho_c}
\end{equation}
where $s_0=2.8912\times 10^9\, {\rm m}^{-3}$ is the entropy density today, $\rho_c=10.537\, h^2\, {\rm GeV\, {\rm m}}^{-3}$ is the critical density and $m_\chi$ the DM mass.

\begin{figure}[!th]%
\centering
\includegraphics[height=0.4\textwidth]{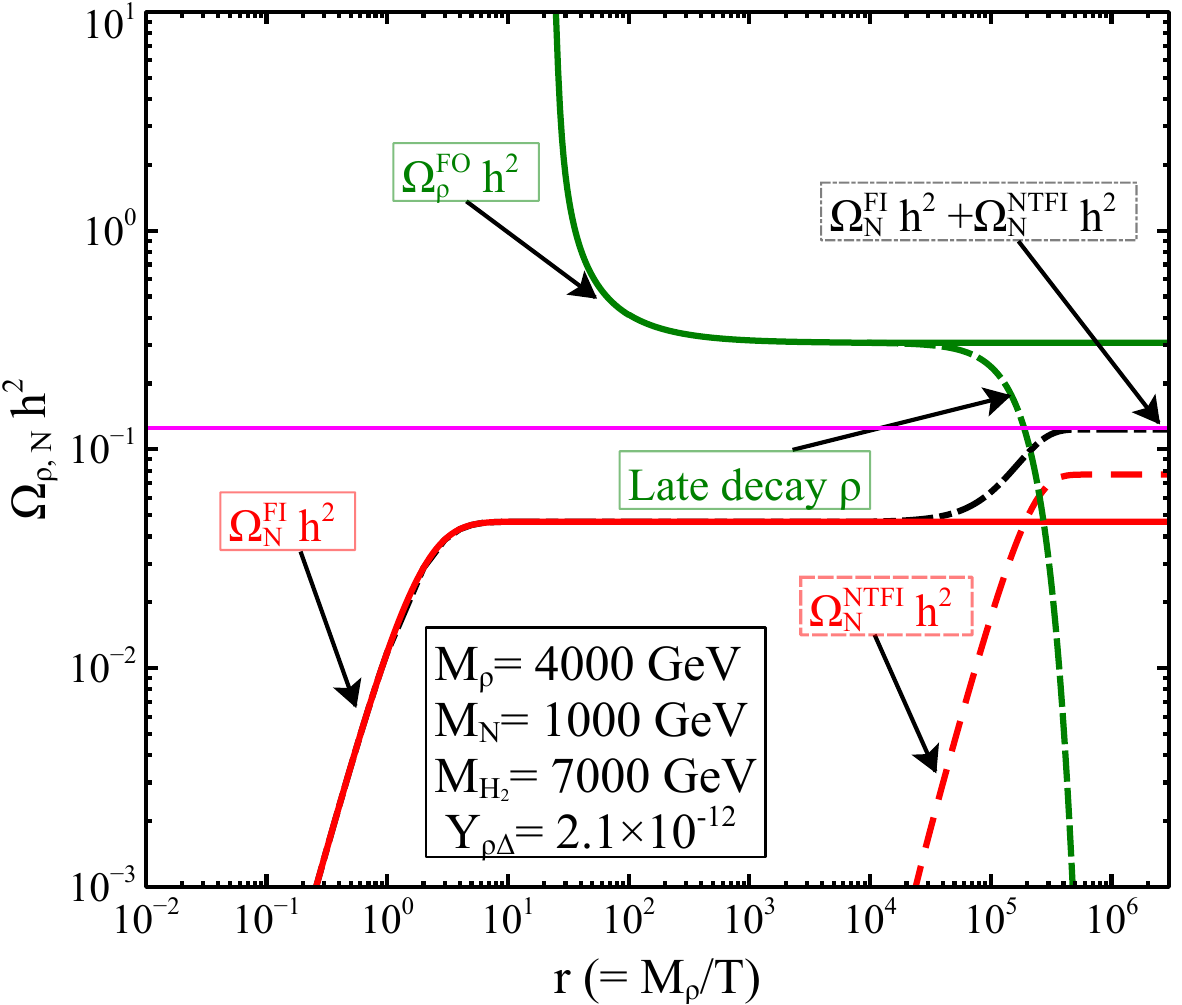}
\includegraphics[height=0.44\textwidth,width=0.5\textwidth]{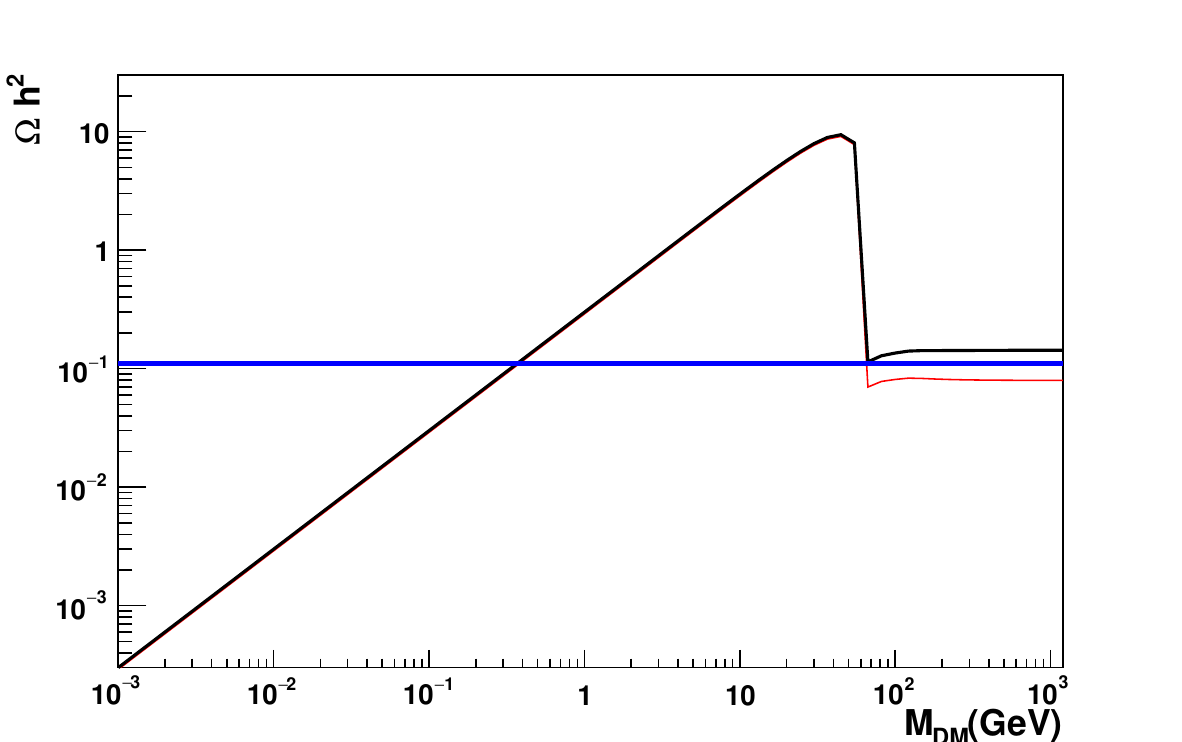}
\caption{Left: Production of DM in the singlet/triplet fermionic  model with the contributions from thermal and non-thermal freeze-in, see details in~\cite{Belanger:2022gqc}. Right: Relic density in the singlet scalar DM model for $\lambda_{HS}=10^{-11}$ assuming Maxwell-Boltzmann (red) or  Bose-Einstein statistics (black)~\cite{Belanger:2018ccd}}.
\label{fig1}
\end{figure}

For dark sectors containing several new particles, FIMPs can also be produced from the decay of a WIMP($\chi'$) after it freezes-out. This is called the superWIMP or non-thermal freeze-in (NTFI) mechanism and was considered long ago  in SUSY models with gravitinos~\cite{Feng:2003xh}. In this mechanism,  the abundance of the FIMP is inherited from that of the WIMP and the total relic density of the FIMP can receive both contributions from thermal and non-thermal  freeze-in ,
\begin{equation}
\Omega^{NTFI}_\chi h^2= \frac{M_\chi}{M_{\chi'}} \Omega_{\chi'}^{FO} h^2 \;\;\;;\;\;   \Omega_\chi h^2= \Omega^{NTFI}_\chi h^2 + \Omega^{FI}_\chi h^2
\end{equation}
where $\Omega_{\chi'}^{FO}h^2$ is the relic density of the WIMP obtained through the freeze-out mechanism. 
In this general picture, one must solve two coupled Boltzmann equations for $\chi$ and $\chi'$. \cref{fig1} shows an example where decay processes  before and after FO both contribute to the total relic density. Here the model is the SM extended with triplet fermions ($\rho$) whose neutral component is a WIMP, a singlet fermion ($N$) which acts as the FIMP as well as a triplet scalar.  The small mixing of  the scalar triplet  with the SM scalar doublet leads to the late decay process $\rho\to N H_1$ that mainly contributes to DM formation after the FO of $\rho$. The decay $H_2\to N \rho$ gives the main contribution to the production of $N$ at early times, while the decay $\rho\to N H_1$ which is suppressed by the small mixing angle between  the scalar triplet  with the SM scalar doublet,  takes place mostly after the FO of $\rho$.

Alternatively, it may be that the WIMP rather than the FIMP is the lightest particle of the dark sector, hence the WIMP is the DM. Nevertheless the FIMP can contribute to DM production through its late decay into the WIMP after it freezes out, thus increasing the relic density of the WIMP~\cite{Hall:2009bx,Belanger:2022gqc}. In these scenarios the DM signatures would then correspond to the usual search for WIMPs although the WIMP annihilation can be larger than in the standard freeze-out scenario thus leading to a boost factor for indirect detection. 

\subsection{Statistics, Thermal masses, symmetry breaking}
In the freeze-in framework,  DM production starts at high temperatures, this entails a dependence on initial conditions and potentially on the details of reheating as well as on several other effects that can become important, namely, quantum statistics, finite temperature corrections as well as phase transition and symmetry breaking effects. 

It is customary to assume that the phase space distribution of bath particles follows a Maxwell-Boltzmann(MB) distribution, a proper treatment of freeze-in however may require 
 to take into account  the Fermi-Dirac or Bose-Einstein phase space distribution of bath particles. In Ref.~\cite{Belanger:2018ccd} it was shown that the inclusion of quantum statistics effects could  lead to a factor two variation in the predicted value of the relic density especially when  driven by annihilation processes initiated by bosons.  The effect is  smaller in the case of fermions and especially  when decays are the dominant processes. 
\cref{fig1} (right)  illustrates  the difference in the relic density using Maxwell-Boltzmann and Bose-Einstein (BE) distributions in the case of the scalar singlet dark matter ($S$) model. Here the coupling is taken  to be $\lambda_{hS}v/\sqrt{2}\, h S^2$ with $\lambda_{hS}=10^{-11}$ . At low masses the decay process $h\to SS$ dominates DM formation and MB approximates well the exact result, the measured value for the relic density is obtained for $M_{DM}\approx 0.3\,\rm{GeV}$. When this process is no longer kinematically accessible, it is rather the annihilation processes through gauge bosons pair annihilation that become dominant, here we see a large increase in the relic density with the BE distribution. Note that the impact of quantum statistics effects is somewhat tamed when including  finite temperature effects~\cite{Bringmann:2021sth}.

 At the high temperatures in the Early Universe where freeze-in starts, interactions with the plasma modify interactions among particles. 
 In the hard thermal loop approximation, it was argued in Ref.~\cite{Weldon:1982aq,Weldon:1989ys,Braaten:1989mz} that these effects can be taken into account by replacing the pole mass  by the Debye mass, which is ${\cal O}(\alpha_s T^2)$ or ${\cal O}(\alpha_W T^2)$ for strongly or weakly interacting particles respectively and by renormalizing all  couplings at the scale $\mu=2\pi T$ using renormalization group equation in vacuum.  Introducing a thermal mass can regulate the enhancement in the forward direction for $t$-channel processes and/or can open up new annihilation or decay channels, for example new decay processes involving  the massive photon or  gluon~\cite{Dvorkin:2019zdi,Belanger:2020npe}. These effects can strongly impact the relic density.

Another effect of DM production at high temperature is that it starts in a regime where electroweak symmetry is not broken. This is typical in models where DM connects to SM via Higgs portal. Apart from the fact that DM production channels differ substantially before and after electroweak symmetry breaking (EWSB) and therefore gives rise to different phenomenology~\cite{Bhattacharya:2021rwh}, if the DM is a scalar ($S$), there is an additional production during the electroweak phase transition, when the temperature is very close to the critical temperature $\approx$ 160 GeV. This additional production arises from the temperature dependent mixing angle, which becomes substantially enhanced for a short period of time right after EWPT and causes rapid conversion of Higgs boson into $S$ through oscillations~\cite{Heeba:2018wtf}.

\subsection{Beyond standard FIMP production}

In models with an extended dark sector, several phases of out-of-equilibrium DM production can occur, depending on the interaction strength within the dark sector. One such phase is the sequential freeze-in as proposed in Ref.~\cite{Hambye:2019dwd}, which consists of the out-of-equilibrium production of mediators from SM particles followed by DM production from non-thermalized mediator annihilation/decay. It occurs when the very small coupling of the mediator to the SM prevents it from reaching equilibrium with the bath. 
To be concrete let us consider a simplified model where the SM is extended with a light scalar mediator, $\phi$,  and a Dirac fermion, $\chi$, the latter being the DM candidate~\cite{Belanger:2020npe},
\begin{equation}
{\cal L}_{int}= y_\chi\, \phi \bar\chi\chi +y_q\, \phi \bar{q}q\;.
\end{equation}
Depending on the size of the couplings there are three DM production modes, 1)  $q\bar{q} \to \chi\chi$, 2)  $\phi\phi \to \chi\chi$ where $\phi$ is in equilibrium with the thermal bath and 3)  $\phi\phi \to \chi\chi$ where $\phi$ is out-of-equilibrium. 
The first two cases are straightforward and lead to a relic density proportional to $y_q^2 y_\chi^2$  and $y_\chi^4$ respectively. 
However in the third case where  $\phi$ is not in equilibrium, its phase space distribution $f_\phi(p)$ does not follow a thermal distribution and one needs to first solve an unintegrated Boltzmann equation for the production of $\phi$ before solving for DM production induced by $\phi\phi$ annihilation. 
Here the  thermal masses of quarks, gluons and photons play an important role in mediator production~\cite{Belanger:2020npe}. 
For further discussion on FI from out of equilibrium particles, see also Ref.~\cite{Aboubrahim:2023yag}.

\section{Signatures of FIMPs }
\subsection{Colliders}

Typically the couplings of FIMPs are too small for these particles to be directly pair produced at a collider. However they can be produced from the decay of a heavier particle, Y,  which is in thermal equilibrium with the thermal bath. At the LHC,  Y could be copiously produced and  because of the feeble coupling would be long-lived. Thus the most relevant searches are those for LLPs, either stable at the collider scale or leading to displaced vertices. For the standard freeze-in scenario, the lifetime of the particle Y with a mass at the weak scale is greater than $c\tau\approx 10^4\, \rm{m}$, thus the most relevant searches are those for heavy stable charged particles (HSCP) when Y is charged. 
When Y is neutral and long-lived its invisible decays can lead to the usual MET signatures at the LHC, while the  visible decay modes will be best probed in planned  detectors located some distance from the interaction point such as MATHUSLA~\cite{No:2019gvl}.  

\begin{figure}[!th]%
\centering
\includegraphics[width=0.7\textwidth]{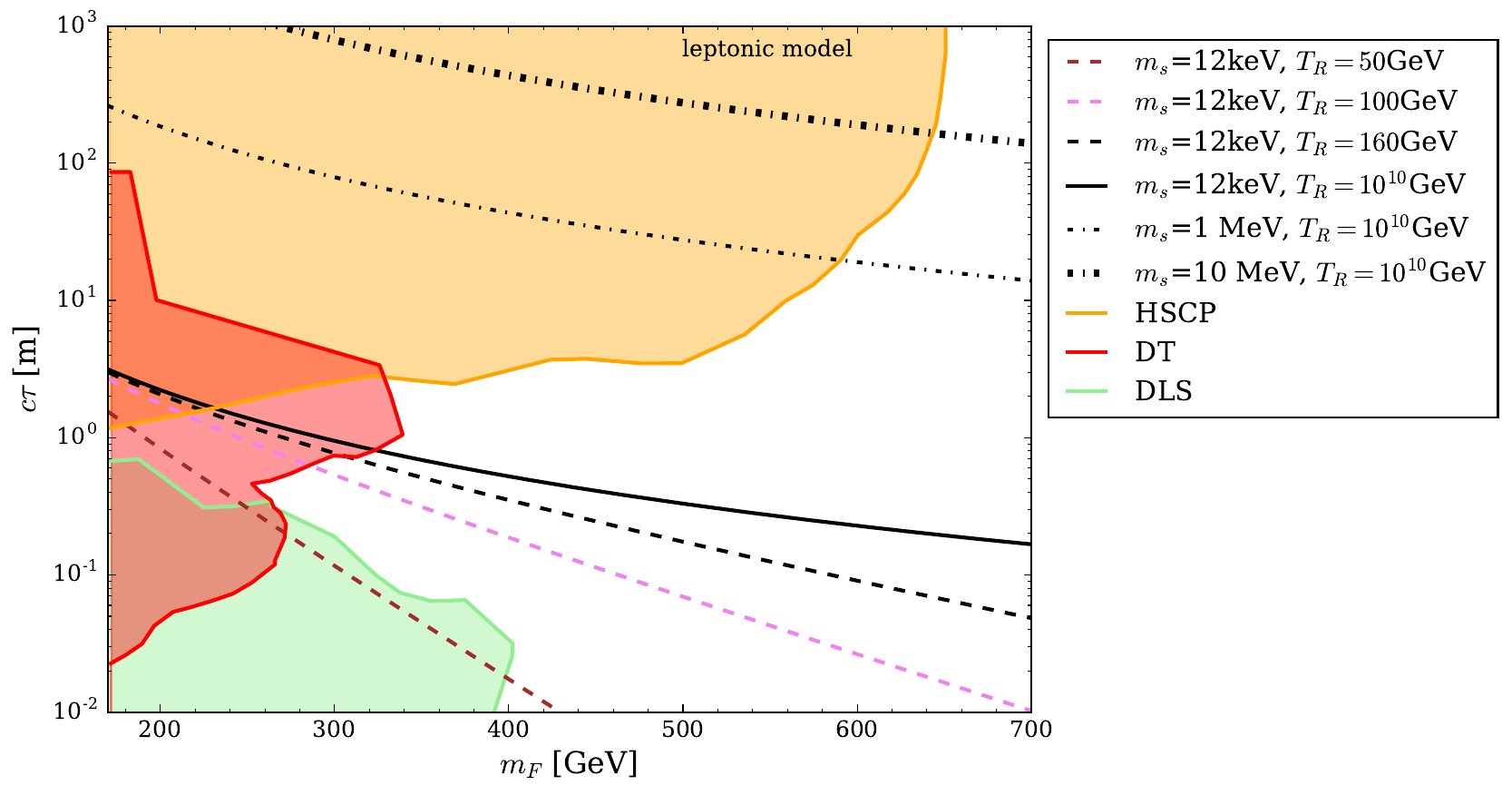}
\caption{LHC constraints for the lepton-like simplified FIMP scenarios. The lines correspond to contours $\Omega_sh^2=0.12$ for the values of $m_s$ and $T_R$ given in the legend~\cite{Belanger:2018sti}.}
\label{fig2}
\end{figure}

Several scenarios have discussed the possibility of larger couplings and thus shorter lifetime within the freeze-in context leading to interesting displaced vertices signatures. These include DM production during the matter dominated era~\cite{Co:2015pka,Evans:2016zau},  lowering the mass of the DM~\cite{Hessler:2016kwm} and/or lowering the reheating temperature~\cite{Belanger:2018sti}.   A concrete example of a FI model which can be probed via  LLP signatures at the LHC was considered in~\cite{Belanger:2018sti}. This simplified model includes a singlet scalar DM and a vector like fermion that couples to an ordinary fermion and DM with a Yukawa coupling. In the case that the vector fermion couples only to leptons, the impact of varying the reheating temperature and the DM mass  on the LHC limits for HSCP, disappearing tracks (DT) and displaced leptons (DLS) is displayed in \cref{fig2}. 

 For light DM (GeV scale or below) an extensive search programme for feebly coupled particles is being put in place at the high energy and high intensity frontier,  see~\cite{Antel:2023hkf} for a comprehensive presentation. While it is generally   challenging to probe the small  couplings relevant for  FIMPs, in specific scenarios - for example with a right-handed neutrino -  
 it was shown that the FI parameter space can be probed in beam-dump or fixed target experiments~\cite{Barman:2022scg}. In addition, a  light mediator  (with a mass $\lesssim 10\, \rm{GeV}$)  can potentially be probed in the beam-dump experiments dedicated to the decays of rare mesons  provided its coupling to quarks is not too suppressed~\cite{Belanger:2020npe}.

\subsection{Direct detection}

Considering the small couplings involved, one might think that  looking for FIMPs via direct detection of its interaction with nuclei in a large detector would be fruitless.
However the small coupling can be somewhat compensated by having a light mediator. The rate for direct detection on nuclei~\cite{Belanger:2008sj,Belanger:2020gnr} is given by, 
\begin{equation}
\frac{dN}{dE}=\frac{\rho_0 \bar\sigma_{SI} }{2 m_\chi \mu^2_{\chi N} } M_{det} T F^2(q) I(E) \frac{m_\phi^4}{(q^2+m_\phi^2)^2}
\end{equation}
where $\bar\sigma_{SI}\propto 1/m_\phi^4$ is the spin independent cross-section at zero momentum transfer, $\mu_{\chi N}$ is the DM-nucleus reduced mass, $\rho_0$ the  DM density today, $M_{det}$ the detector mass and $T$ the exposure time.   $F(q)$ is a form factor with $q$ the momentum transfer,  and $I(E)=\int_{v_{min}(E)} f(v)/v dv$ takes into account  the DM velocity distribution. Note that the large enhancement of the cross-section  saturates when the mass of the mediator becomes of the order of $q^2=2E M_N$, which is typically 100 MeV for a weak scale DM. Ref.~\cite{Hambye:2018dpi,Essig:2015cda} showed that  freeze-in could be probed for light mediators.

For DM at the GeV scale or below it was suggested to exploit DM scattering on electrons. 
Studies have shown that future detectors can cover the favored freeze-in region in the case of a light vector mediator for example~\cite{Essig:2017kqs,XENON:2021qze,Emken:2019tni}.

\subsection{Cosmological aspects}

The presence of FIMPs, which often leads to LLPs, can impact the successful predictions of Big Bang Nucleosynthesis on the abundance of light elements. 
Indeed the decay of LLPs can inject energetic particles that can trigger non-thermal nuclear processes. For particles with lifetime ranging from $0.1-10^4\,$s the relevant bounds are due to the hadronic part of the decay which can alter the $n/p$ ratio thus shifting the predictions for $He^4$ and $D/H$. 
The key elements  are the hadronic branching ratio of the LLP, $B_{had}$, the net energy carried away by hadrons, $E_{vis}$, the yield of the LLP, $Y_{LLP}$  and the lifetime of the LLP, $\tau_{LLP}$. 
Generic constraints in the plane $B_{had} E_{vis} Y_{LLP}$ vs $\tau_{LLP}$ using both the $D/H$ ratio -- relevant for $\tau>100$s -- and the abundance of $He^4$ for shorter lifetimes  were derived in Ref.~\cite{Kawasaki:2004qu,Kawasaki:2015yya}. These have been shown to put strong constraints on models with FIMP DM and a LLP, for example in the triplet/singlet fermionic model~\cite{Belanger:2022gqc} and in an inert doublet and singlet model~\cite{Belanger:2021lwd}. 
          
LLPs with longer    lifetimes                                       can also impact the CMB leading to spectral distorsions as well as to anisotropies in the power spectra.  The former provides constraints on the amount of electromagnetic energy injected before $10^{12}$s while the latter constrains particles with lifetimes $\tau>10^{12}\,$s~\cite{Poulin:2016anj,Fixsen:1996nj}.

 Light DM also contributes to the cooling of stars. The observations of globular clusters severely constrain DM lighter than 100 keV, in particular for the parameter space preferred by freeze-in~\cite{Chang:2019xva}.

\section{Conclusion}
In summary, different mechanisms can contribute to feeble particle production in the early Universe, and contrary to freeze-out this mechanism may depend on the initial conditions and on the reheating temperature $T_R$. Moreover thermal effects can be important. Despite the very small couplings of the FIMP, various signatures of FIMPs can be detectable at the high energy and/or high intensity frontier, including  direct searches of LLPs or indirect searches for the mediator. Moreover FIMPs can be probed via direct detection if the mediator is light and with cosmology, in particular through effect on BBN and/or distorsions of the CMB.

\acknowledgments
This work was funded in part by the Indo-French Centre for the Promotion of Advanced
Research (Project title: Beyond Standard Model Physics with Neutrino and Dark Matter
at Energy, Intensity and Cosmic Frontiers, Grant no: 6304-2). SC is supported by the UKRI Future Leaders Fellowship DARKMAP.
The work of AP was carried out within the scientific program “Particle Physics and Cosmology” of the Russian National Center for Physics and Mathematics.

\bibliography{biblio}
\end{document}